\documentclass[lettersize,journal]{IEEEtran}
\usepackage{amsmath,amsfonts}
\usepackage{array}
\usepackage[caption=false,font=normalsize,labelfont=sf,textfont=sf]{subfig}
\usepackage{textcomp}
\usepackage{stfloats}
\usepackage{url}
\usepackage{verbatim}
\usepackage{graphicx}
\usepackage{cite}
\usepackage[ruled,vlined,linesnumbered]{algorithm2e}
\usepackage{graphicx}   
\usepackage{booktabs}
\usepackage{tikz}
\newcommand{\BugForge}{\textsc{\itshape{BugForge}}\xspace}

\newcommand{\squirrelU}{\squirrel{}$^{U}$\xspace}
\newcommand{\griffinU}{\griffin{}$^{U}$\xspace}
\newcommand{\squirrelB}{\squirrel{}$^{B}$\xspace}
\newcommand{\griffinB}{\griffin{}$^{B}$\xspace}
\newcommand{\squirrelUB}{\squirrel{}$^{U+B}$\xspace}
\newcommand{\griffinUB}{\griffin{}$^{U+B}$\xspace}

\newif\ifshowtoAF
\showtoAFtrue  
\newif\ifshowtodos
\showtodostrue

\newcommand{\totalrawpoc}{35,530\xspace}
\newcommand{\totalBugs}{35\xspace}
\newcommand{\totalConfirmedBugs}{22\xspace}
\newcommand{\fuzzingBugs}{23\xspace}
\newcommand{\regressionBugs}{9\xspace}
\newcommand{\crossDBMSBugs}{3\xspace}

\newcommand{\totalreport}{37,632\xspace}
\newcommand{\mysqlreport}{18,212\xspace}
\newcommand{\mariadbreport}{12,442\xspace}
\newcommand{\pgreport}{4,163\xspace}
\newcommand{\monetdbreport}{2,815\xspace}

\newcommand{\mysqlexecutabletestcases}{16,112\xspace}
\newcommand{\mariadbexecutabletestcases}{11,426\xspace}
\newcommand{\pgexecutabletestcases}{3,616\xspace}
\newcommand{\monetdbexecutabletestcases}{2,444\xspace}

\newcommand{\timerange}{1998 to 2026\xspace}

\newcommand{\mysqlbug}{11\xspace}
\newcommand{\mariadbbug}{11\xspace}
\newcommand{\pgbug}{3\xspace}
\newcommand{\monetdbbug}{10\xspace}

\newcommand{\squirrel}{\textsc{Squirrel}\xspace}
\newcommand{\griffin}{\textsc{Griffin}\xspace}

\newcommand{\mysql}{MySQL\xspace}

\newcommand{\pg}{PostgreSQL\xspace}
\newcommand{\mariadb}{MariaDB\xspace}

\newcommand{\monetdb}{MonetDB\xspace}

\newtheorem{exampleenv}{Example}

\newcommand{\mytilde}{\raise.17ex\hbox{$\scriptstyle\mathtt{\sim}$}}

\usepackage{enumitem}
\usepackage{xspace}
\usepackage{listings}
\usepackage{color} 
\usepackage{fontawesome}
\usepackage{pifont}
\usepackage{float} 
\usepackage{textcomp}
\usepackage{algpseudocode}
\usepackage{multirow}
\usepackage{booktabs}   
\usepackage{caption}    
\usepackage{tabularx}   
\usepackage{ragged2e}   
\usepackage{csquotes}

\lstdefinestyle{sqlstyle}{
    language=SQL,
    basicstyle=\ttfamily\footnotesize,
    frame=single,
    xleftmargin=0.1cm,
    xrightmargin=0.1cm,
    rulecolor=\color{gray},
    commentstyle=\color{darkgray},
    commentstyle=\color{gray},
    numberstyle=\tiny\color{darkgray},
    numbersep=5pt,
    tabsize=4,
    breaklines=True,
    breakatwhitespace=false,
    showspaces=false,
    showstringspaces=false,
    showtabs=false,
    upquote=true,
    columns=fullflexible,
    escapeinside={(*@}{@*)}    
}

\begin{document}

\title{\BugForge: Constructing and Utilizing DBMS Bug Repository to Enhance DBMS Testing
}

\author{Dawei~Li \textsuperscript{\dag}, Qifan~Liu \textsuperscript{\dag}, Yuxiao~Guo, Jie~Liang, Zhiyong~Wu,\\
Chi~Zhang, Jingzhou~Fu, Haogang~Mao, Zhenyu~Guan, and Yu~Jiang
\thanks{Dawei Li, Qifan Liu, Yuxiao Guo, Jie Liang, Haogang Mao, and Zhenyu Guan are with Beihang University, Beijing, China
(e-mail: lidawei@buaa.edu.cn; imchifan@163.com; yuxiaoguo@buaa.edu.cn; liangjie.mailbox.cn@gmail.com; maohaogang@126.com; guanzhenyu@buaa.edu.cn). Corresponding author: Jie Liang (liangjie.mailbox.cn@gmail.com).}%
\thanks{Zhiyong Wu, Chi Zhang, Jingzhou Fu, and Yu Jiang are with Tsinghua University, Beijing, China
(e-mail: wzy199936@163.com; chi-zhang@mail.tsinghua.edu.cn; fuboat@outlook.com; jiangyu198964@126.com).

}%
}


\markboth{Journal of \LaTeX\ Class Files,~Vol.~14, No.~8, August~2021}%
{Shell \MakeLowercase{\textit{et al.}}: A Sample Article Using IEEEtran.cls for IEEE Journals}

\IEEEpubid{0000--0000/00\$00.00~\copyright~2021 IEEE}

\maketitle

\begingroup
\renewcommand\thefootnote{\dag}
\footnotetext{Dawei Li and Qifan Liu contributed equally to this work.}
\endgroup

\begin{abstract}
DBMSs are complex systems prone to bugs that may lead to system failures or compromise data integrity.
Establishing unified DBMS bug repositories is crucial for systematically organizing bug-related data, enabling code improvement, and supporting automated testing.
In particular, bug reports often contain valuable test inputs and bug-triggering clues that help explore rare execution paths and expose critical buggy behavior, thereby guiding automated DBMS testing. 
However, the heterogeneity of bug reports, along with their incomplete or inaccurate content, makes it challenging to build unified repositories and convert them into high-quality test cases.

In this paper, we propose \BugForge, a framework that constructs standardized DBMS bug repositories and leverages them to generate high-quality test cases to enhance  DBMS testing.
Specifically, \BugForge progressively collects bug reports, then employs syntax-aware processing and input-adaptive raw PoC extraction to construct a DBMS bug repository. 
The repository stores structured bug-related data, including bug metadata and raw PoCs that entail potential bug-triggering semantics.
These data are further refined into high-quality test cases through semantic-guided adaptation, thereby enabling enhanced DBMS testing methods, including DBMS fuzzing, regression testing, and cross-DBMS bug discovery.
We implemented \BugForge for \pg{}, \mysql{}, \mariadb{}, and \monetdb{}, totally integrated \totalreport bug reports spanning up to 28 years.
Based on the repository, \BugForge uncovered \totalBugs{} previously unknown bugs with \totalConfirmedBugs confirmed by developers,
demonstrating the value of constructing and utilizing bug repositories for DBMS testing.

\end{abstract}


\section{Introduction}
Database Management Systems (DBMSs) are integral to modern software applications, where their correctness and reliability are paramount to the stability and data consistency of overlying applications~\cite{DBMS}. 
Nevertheless, the inherent complexity and vast codebases of DBMSs, which feature multiple, tightly coupled subsystems such as query optimization, fault recovery, and storage management, render them susceptible to subtle bugs~\cite{pgbugs, mysqlbugs}. 
Once exploited, these bugs can result in severe consequences, including data breaches, service interruptions, or system crashes.

To effectively manage and analyze DBMS bugs, \textbf{establishing unified DBMS bug repositories} is crucial in industry practice. 
Such repositories, which include bug-triggering information, provide practical value for both code maintenance and testing.
Developers can use the repository to support daily maintenance, track upstream fixes, and issue security alerts to downstream vendors, while test engineers can use this information as seeds for fuzzing campaigns, as robust regression test cases, and as standard inputs for database evaluation.
Moreover, the structured records facilitate fault localization, program repair, and defect prediction by making recurring patterns and common error triggers easier to detect.



Practically, both DBMS vendors and testing researchers have developed mechanisms to manage DBMS bugs.
Database vendors typically collect, verify, and categorize user-submitted issues through official bug reporting platforms (e.g., \mysql{} Bug System~\cite{mysqlbugs} and \pg{} Bug Tracker~\cite{pgbugs}).
Test researchers, on the other hand, often disclose bugs in public communities and further organize them into structured datasets for analysis, testing, and methodology validation~\cite{manueldbbugs, fubuglist, jinbuglist}. 
These efforts improve the transparency and security responsiveness within the DBMS ecosystem.

%

\IEEEpubidadjcol

\textbf{However, existing DBMS bug management remains incomplete and fragmented, limiting the full utilization of bug-related data.}
First, bug reports are often incomplete or ambiguous, since many are reported informally by users, making it difficult for developers or downstream vendors to determine whether issues are true faults or false positives.
Second, most repositories focus on logging bugs rather than enabling structured analysis or automated processing. 
The lack of standardized formats and interfaces, combined with manual maintenance, restricts systematic use of bug information in reports.
For example, many bug reports include proof-of-concept (PoC) code snippets that may contain syntax or semantic errors, making them unusable without adaptation. We refer to these as \textbf{raw PoCs}, defined as the original PoC contents provided in bug reports before any repair, completion, or validation.
Third, bug information is siloed in vendor-specific repositories, making it difficult to leverage insights from one report to analyze and identify similar bugs in related DBMSs. 
These limitations hinder bug reports from serving as reliable resources for testing. We highlight two key challenges in leveraging DBMS bug reports for automated testing.

(1) \textit{Mining bug information from unstructured and evolving DBMS bug reports that interleave SQL semantics with natural-language descriptions.}
Most reports are written in unstructured natural language; for example, raw PoCs are often embedded in descriptive text rather than presented as clear code blocks, which makes them difficult to automated extract.
Moreover, DBMS bug reports come from multiple sources, including vendor platforms, community forums, and research submissions. The content of a report may evolve throughout its lifecycle, such as status changes or developer feedback.
(2) \textit{Restoring executability and preserving semantic richness to construct high-quality test cases that depend on specific database states}:
Even when raw PoCs can be extracted, they often lack configuration details or contain syntactic and semantic errors. Attempting to execute them directly often fails, making it difficult to construct high-quality test cases for DBMS testing. In this paper, \textit{high-quality test cases are characterized by both reliable executability and preservation of the semantic richness carried by raw PoCs, especially those bug-triggering cases.} This requires \BugForge to infer missing environment and state information, correct execution-blocking issues, and preserve the semantic richness.

In this paper, we propose \BugForge
, an end-to-end framework for constructing and utilizing DBMS bug repositories from real-world reports, particularly enabling the effective generation of high-quality cases from raw PoCs for DBMS testing.
(1)~To address the first challenge, \BugForge employs an automated repository construction pipeline. It first performs incremental report collection and syntax-aware heuristic processing to filter redundant noise and identify PoC-related fragments from heterogeneous report content, and then applies an input-adaptive LLM framework with RAG to extract raw PoCs.
(2)~To address the second challenge, \BugForge introduces a semantic-guided adaptation strategy. It first performs feedback-driven adaptation to repair incomplete or erroneous inputs, and then applies semantic-constrained adaptation to maintain semantic richness of the raw PoCs. During adaptation, \BugForge also monitors environmental side effects and performs cleanup when necessary to ensure adaptation independence and stability.
(3)~The resulting repository constructed by \BugForge is leveraged to enhance DBMS testing, serving as initial seeds for DBMS fuzzing, supporting regression tests, and enabling cross-DBMS bug discovery.

We implemented \BugForge{} and constructed a comprehensive DBMS bug repository encompassing \pg{}, \mysql{}, \mariadb{}, and \monetdb{}. The reports originate from diverse sources, including official vendor platforms and community submissions. 
The repository consolidates \totalreport DBMS bug reports collected from 1998 to 2026.
We also utilized \BugForge{} for testing in three aspects: DBMS fuzzing, regression testing, and cross-DBMS testing.
The results revealed that \BugForge{} helped find {\totalBugs} bugs in four popular DBMSs. 
Ultimately, we reported {\totalBugs} bugs, of which {\fuzzingBugs} were found by DBMS fuzzing, {\regressionBugs} were regression bugs, and {\crossDBMSBugs} were found by cross-DBMS testing.
Out of these, {\totalConfirmedBugs} bugs have been confirmed. One of the \pg developers praised one of our reported bugs as an interesting example.
In summary, our paper makes the following contributions:

\begin{itemize}[leftmargin=2em]
\item We find that unified DBMS bug repositories are still lacking, and building them is challenging due to the unstructured and incomplete bug reports. Moreover, these reports are difficult to convert into high-quality test cases for automated DBMS testing.


\item We propose \BugForge{}, an automated framework for constructing and leveraging DBMS bug repositories. It performs report collection and syntax-aware processing to construct a repository, then generates high-quality test cases by semantic-guided adaptation. The resulting cases can support automated DBMS testing like fuzzing.

\item 
We use \BugForge{} to build a DBMS bug repository containing {\totalreport} bug reports across 4 popular DBMSs like \pg{}. Leveraging it, we discover \totalBugs previously unknown bugs, among which \totalConfirmedBugs have been confirmed.

\end{itemize}

\section{Background and Motivation}

\begin{figure}[!b]
    \centering
    \includegraphics[scale=0.5]{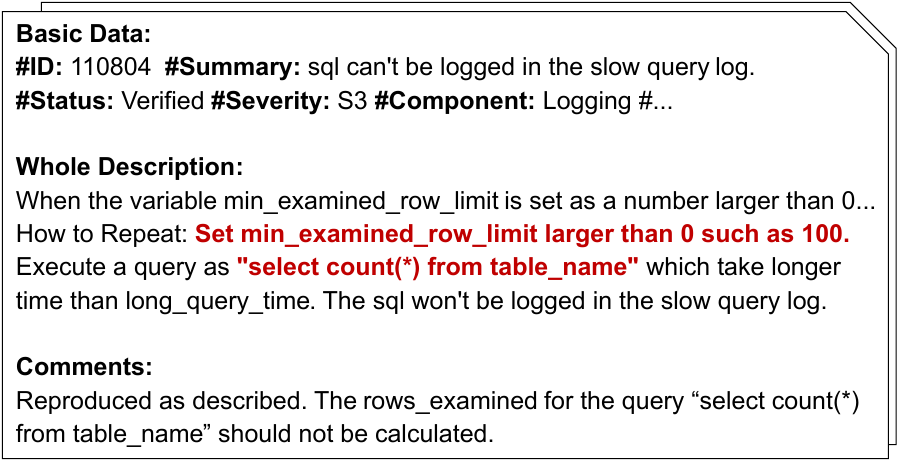}
    \caption{\mysql{} Bug Report \texttt{\#110804} contains a raw PoC that is difficult to extract automatically. \mdseries 
    This report complicates automated analysis, as the PoC is conveyed in prose instead of being explicitly presented in a distinct code block.
    The sequence of statements to reproduce the bug, must be manually reconstructed from the content of bug report.
    }
    \label{fig:bug_example}
\end{figure}

\textbf{Value of Bug Repository in DBMS Testing.}
A \textit{bug repository} generally refers to a curated collection of buggy programs, providing a valuable platform for software testing, automated repair, and model training.
While many critical domains already maintain mature bug repositories~\cite{10.1145/2610384.2628055, 10.1145/1321631.1321702}, DBMSs still lack such a resource. 
As shown in Figure~\ref{fig:bug_example}, a \textit{bug report} serves as the basic entry or record within a bug repository, containing essential information about a specific bug.
Typical fields in a bug report include a natural language description of the problem, version, and environment information, logs, and typically an attached PoC.

A PoC is a minimal artifact that demonstrates a bug or vulnerability. When available, it is the most valuable part of a report because it converts a textual description into a usable test case. Within DBMSs, PoCs typically consist of sequences of SQL statements that expose crashes, inconsistencies, or incorrect query results. These statements may contain syntax, semantic, or configuration errors and cannot always be executed directly.
Therefore, we refer to such unprocessed code as \textbf{raw PoC}.
By systematically collecting and organizing these bug reports and raw PoCs, unified bug repositories enable more effective automated testing, regression analysis, and overall maintenance of DBMS software. 

\textbf{Challenges in Constructing and Utilizing DBMS Bug Repository.}
Despite the rich information embedded in bug repositories and their potential applications in areas such as testing, effectively utilizing them remains challenging:

(1) First, bug reports are typically written in unstructured natural language, making automated extraction of complete SQL sequences difficult. 
For example, Figure~\ref{fig:bug_example} shows a \mysql{} bug report that contains a raw PoC that is difficult to extract automatically. 
Unlike standard code blocks, this PoC is embedded in prose, requiring the sequence of SQL statements \texttt{SET min\_examined\_row\_limit =100} and \texttt{SELECT count(*) FROM table\_name} to be manually inferred and reconstructed. 
Such unstructured formats hinder the extraction and reuse of raw PoCs, exacerbating the challenges of automated bug information mining.

\begin{figure}[htbp]
    \centering\includegraphics[width=1\linewidth]{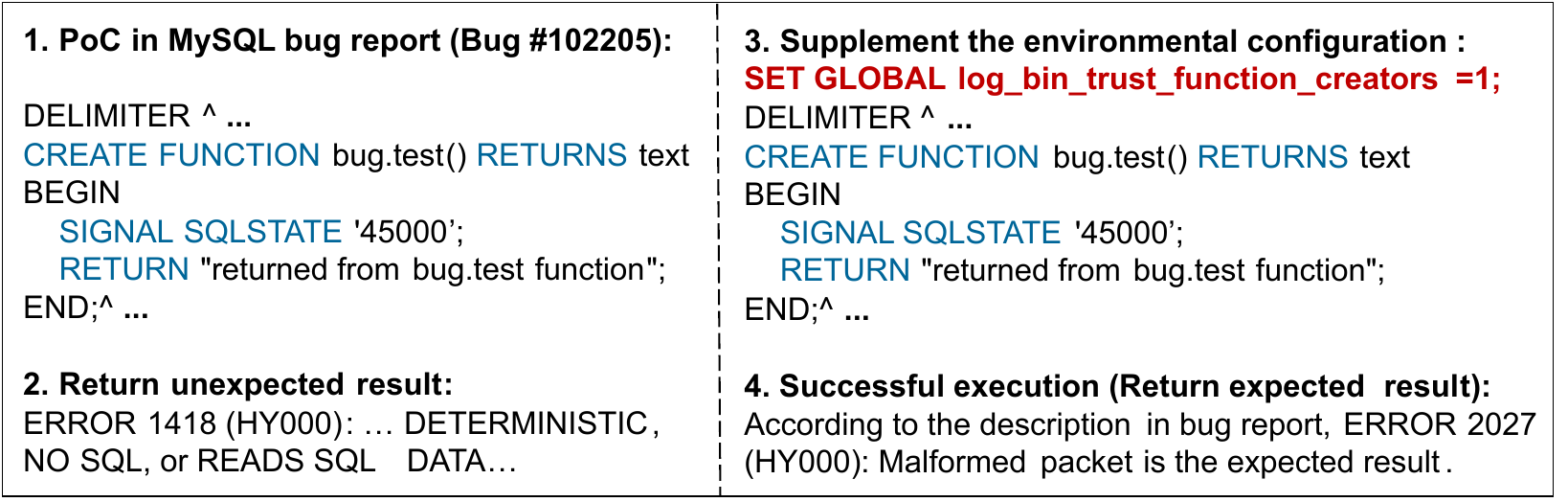}
    \caption{\mysql{} Bug Report \texttt{\#102205} misses the configuration  required for successful execution. \mdseries The bug is unreproducible because the bug report lacks the configuration information for the \texttt{log\_bin\_trust\_function\_creators} system variable, but it triggers the expected result after \BugForge adapts the raw PoC.
    }
    \label{fig:motivation}
\end{figure}

(2) Second, raw PoCs in bug reports are sometimes incomplete, inconsistent, or tightly coupled with specific runtime environments. 
PoCs derived from DBMS bug reports frequently lack configuration details and may contain syntactic or semantic errors, which prevent them from being utilized for DBMS testing.
Successful adaptation requires completing missing configuration details, correcting code errors, and preserving semantic richness with the original raw PoC.
For example, in Figure~\ref{fig:motivation}, the raw PoC extracted from \mysql{} Bug Report \texttt{\#102205} \cite{mysql102205}, when executed, fails not by exposing the intended bugs, but by producing a preliminary \textquotesingle ERROR 1418\textquotesingle, which stems from a missing configuration and halts any further analysis or utilization. 
This scenario represents a critical bottleneck where valuable bug data is effectively lost, stymying automated testing efforts.

These issues mean that although DBMS bug reports contain valuable information, their raw form is often unsuitable for automated testing pipelines. Many raw PoCs cannot be utilized effectively, hindering systematic reproduction, regression testing, fuzzing campaigns, and cross-database evaluations. Addressing these challenges requires methods for extracting, adapting, and standardizing PoCs from heterogeneous and partially incomplete DBMS bug reports.

\textbf{Basic Idea of \BugForge.}
The basic idea of \BugForge{} is to use syntax-aware processing to extract raw PoCs and transform them into high-quality test cases through semantic-guided adaptation.
(1) To mine bug information and extract raw PoCs, \BugForge first employs syntax-aware heuristic analysis to filter redundant noise, structure bug reports, and derive raw PoC-related fragments. Then \BugForge leverages an input-adaptive LLM with retrieval-augmented generation to identify the exact SQL statements and code fragments associated with bug triggering, thereby extracting raw PoCs from the fragments and constructing the DBMS bug repository.
(2) To generate high-quality test cases for enhancing DBMS testing, \BugForge{} adopts a semantic-guided adaptation strategy for raw PoCs. 
It first performs feedback-driven adaptation to infer missing configurations and fix syntactic or semantic errors, and then applies semantic-constrained adaptation to preserve as much semantic richness in raw PoCs as possible. 
Concurrently, \BugForge performs necessary environment cleanup to maintain the stability and independence of adaptation.    

\textbf{Example: A Configuration-Dependent Bug in \mysql{}.}
We illustrate how \BugForge{} can be applied to adapt complex PoCs in \mysql{} through an example of a configuration-dependent bug. 
This bug arises only under specific server settings and results in incorrect query results when certain functions are executed. 

As shown in Figure~\ref{fig:motivation}, \BugForge{} first extracts the raw PoC from Bug Report \texttt{\#102205}.
Then it executes the raw PoC and infers through static analysis from error feedback based on lexical matching that ``ERROR 1418'' is an environmental prerequisite issue.
Leveraging the contextual reasoning of the LLM, \BugForge then autonomously refines the raw PoC, intelligently injecting the necessary command: \texttt{set global log\_bin\_trust\_function\_creators=1}. 
This adaptation allows the execution to proceed past the initial roadblock, successfully reproducing the true underlying bug: ``ERROR 2027 (Malformed packet) and loss connection''. 
After adaptation, the PoC successfully reproduces the bug, allowing developers to validate and investigate the underlying issue. 
These enhanced PoCs can be used as highly effective seeds for fuzzing campaigns to uncover novel bugs or as robust test cases for regression and cross-database testing to ensure system stability and compatibility.


\section{Design}
\begin{figure*}[htbp]
    \centering
    \includegraphics[width=0.97\textwidth]{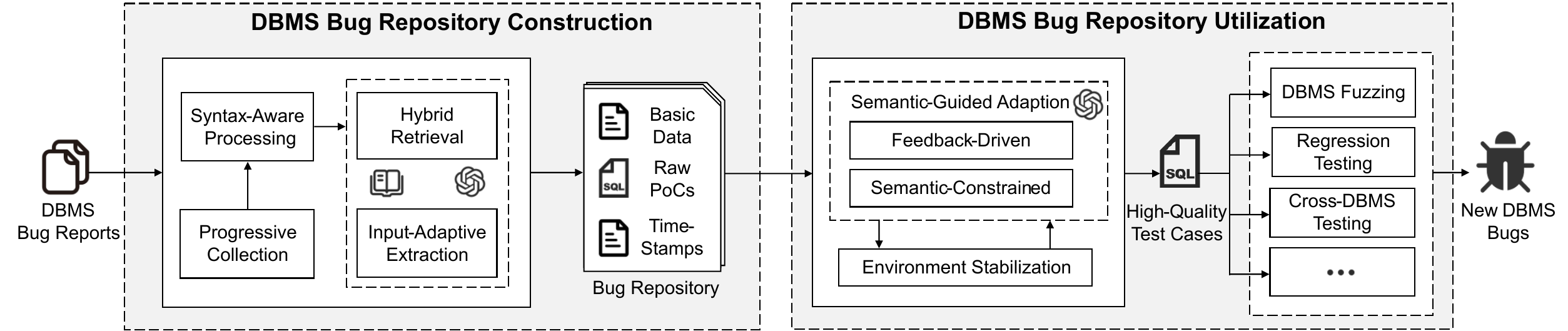}
    \caption{\textbf{The overview design of \BugForge.}
\mdseries 
\BugForge is an end-to-end framework for DBMS bug repository construction and utilization. It progressively collects bug reports, applies syntax-aware processing and the input-adaptive LLM pipeline with RAG to extract raw PoCs and constructs a DBMS bug repository. Next, \BugForge converts raw PoCs into high-quality test cases through semantic-guided adaptation under a stable environment and subsequently employs these cases in DBMS testing (e.g., fuzzing, regression, and cross-DBMS testing).}
    \label{whole_design}
    \vspace{-0.2cm}
\end{figure*}

Figure~\ref{whole_design} illustrates the architecture of \BugForge, which contains two steps: Bug Repository Construction and Bug Repository Utilization.
To construct the repository, \BugForge progressively collects bug reports and combines syntax-aware report processing with input-adaptive LLM mining to extract raw PoCs. In repository utilization, \BugForge applies semantic-guided adaptation to transform raw PoCs into high-quality test cases.
Next, \BugForge uses the high-quality test cases 
to enhance DBMS testing, including DBMS fuzzing, regression testing, and cross-DBMS testing.

\vspace{-0.2cm}
\subsection{Bug Repository Construction }
Bug repository construction aims to collect bug reports from the vast corpus, identify and extract the PoC together with their relevant bug context.
The challenge is that DBMS bug reports are highly unstructured, frequently mixing natural-language discussion, fragmented SQL, execution logs, and configuration details, which makes the automated identification of bug-relevant SQL inputs difficult.

\textbf{Progressive DBMS Bug Reports Collection.}
Bug reports for DBMSs typically originate from diverse sources. To ensure the reliability and accuracy of the collected bugs, we restrict our collection to reports published on official bug tracking platforms, which are publicly accessible and maintained by dedicated developers who rigorously triage, validate, and track reported issues through resolution.
We further filter the collected reports based on their status, retaining only those that have been confirmed or fixed by developers, thereby ensuring that each report corresponds to a real defect acknowledged by the maintainers.
The collected bugs span multiple categories, including system crashes, runtime errors, incorrect outputs, and performance degradation.

Additionally, new reports are constantly being submitted and confirmed, while historical reports may be updated in subsequent periods (e.g., developers replying to reports). In addition, the total number of reports is exceptionally large, repeatedly collecting the entire dataset to capture ongoing updates and revisions is inefficient. To ensure data consistency and timeliness, \BugForge adopts an asynchronous and periodic synchronization strategy. 
It selectively probes existing records based on the time elapsed since their last collection, with adaptive refresh intervals determined by report age. For each report, the system fetches a lightweight upstream snapshot and checks whether key mutable attributes, especially the status and last-modified timestamp, differ from the local copy. Only reports that are newly observed, detected as changed are subjected to full re-collection and local update. This design reduces unnecessary collection overhead while preserving the freshness of the local report repository.

\textbf{Syntax-Aware DBMS Bug Report Processing.}
The precise identification of bug information within bug reports serves as the basis for generating test cases that can be effectively utilized.
Unlike generic software bug reports, DBMS reports often interleave natural-language descriptions with SQL fragments, execution logs, error messages, and configuration details, making traditional manual analysis or pattern-based methods difficult, especially when SQL statements are fragmented, partially omitted, or implicitly embedded in surrounding text. While LLMs offer strong capabilities for report understanding, directly submitting the entire report may overwhelm the model with redundant context and weaken its focus on bug-triggering content. 
Therefore, before raw PoC extraction, \BugForge applies a syntax-aware heuristic analysis pipeline to preprocess DBMS bug reports and locate text content that is potentially indicative of the bug-triggering procedure. We define such content as \emph{PoC-related fragments}: report fragments that may preserve procedural or semantic clues for constructing a raw PoC, but do not necessarily constitute a complete or directly high-quality raw PoC.


\begin{algorithm}[h]
\footnotesize
\label{alg:poc_extraction_simplified}
\KwIn{DBMS bug report: $\mathcal{L}$.}
\KwOut{PoC-related fragments: $\mathcal{P}$, or \textbf{null}.}

\BlankLine
$\mathcal{S} \leftarrow \text{an empty list of (index, content) tuples}$\;
$i \leftarrow 0$\;

\While{$i < |\mathcal{L}|$}{
    \tcp{Capturing Formatted SQL via Text Structure}
    \uIf{$\mathcal{L}[i]$ is a start-of-block tag (e.g., markdown)}{
        $j \leftarrow \text{Find corresponding end tag from } i+1$\;
        \If{$j$ is found}{
            Add $(i, \mathcal{L}[i+1 : j])$ to $\mathcal{S}$; \quad $i \leftarrow j$\;
        }
    }
    \tcp{Capturing Fragmented SQL via SQL Feature}
    \uElseIf{$\mathcal{L}[i]$ is a endpoint (e.g., semicolon)}{
        $j \leftarrow i$; \quad $b\leftarrow 0$\;
        \While{$j \geq 0$}{
            $b \leftarrow b + \text{Count}(\mathcal{L}[j], ')') - \text{Count}(\mathcal{L}[j], '(')$\;
            \lIf{$\text{IsSQLSubject}(\mathcal{L}[j])$ \textbf{and} $b\leq 0$}
            {\textbf{break}} $j \leftarrow j - 1$\;
        }
        Add $(j, \mathcal{L}[j : i+1])$ to $\mathcal{S}$\;
    }
    \tcp{Capturing Implicit SQL via Weight Calculation}
\uElseIf{$\text{CalculateScore}(\mathcal{L}[i]) \geq \theta_{\text{score}}$}{
        Add $(i, [\mathcal{L}[i]])$ to $\mathcal{S}$\;
    }
    $i \leftarrow i + 1$\;
}
\Return $\mathcal{P}$ if $\mathcal{P}$ is not \textbf{null} else \textbf{null}\;
\caption{Syntax-Aware Bug Reports Processing}
\end{algorithm}

Algorithm 1 shows the detailed report process via heuristic analysis.
Given a DBMS bug report $\mathcal{L}$, \BugForge scans $\mathcal{L}$ using a prioritized three-stage strategy that progressively captures SQL code representations ranging from explicitly formatted blocks to fragmented or implicit statements. 
The process first targets explicitly formatted SQL by identifying structural markers, and greedily extracts complete code regions enclosed by standard delimiters such as Markdown backticks (Lines 4–8).
When such formatting is absent, the strategy transitions to recovering fragmented SQL via syntax-aware backtracking. 
Upon encountering a statement terminator (e.g., a semicolon), \BugForge performs a reverse traversal, tracking delimiter balance (e.g., parentheses count $b$) and extending the search backward until a valid SQL starting subject is identified, thereby reconstructing a semantically complete multi-line statement. (Lines 9–16).  
Finally, for the remaining ambiguous lines that may implicitly contain SQL, \BugForge applies a weight-based heuristic to evaluate individual lines against a confidence threshold $\theta$, while down-weighting or filtering redundant content such as debug outputs, thereby retaining compact PoC-related fragments with sufficiently high syntactic probability for subsequent LLM extracting (Lines 17–18).

\textbf{Input-Adaptive Raw PoC Extraction.}
After preprocessing, the PoC-related fragments already capture condensed bug-triggering information from bug reports. 
However, they may still be incomplete, or semantically ambiguous, and thus difficult to directly extract precise raw PoCs. To improve extraction accuracy while controlling context overhead, \BugForge employs an LLM-based extraction framework with hybrid retrieval and adaptive input expansion. 

\begin{figure}[htbp]
\vspace{-0.1cm}
    \centering\includegraphics[width=0.93\linewidth]{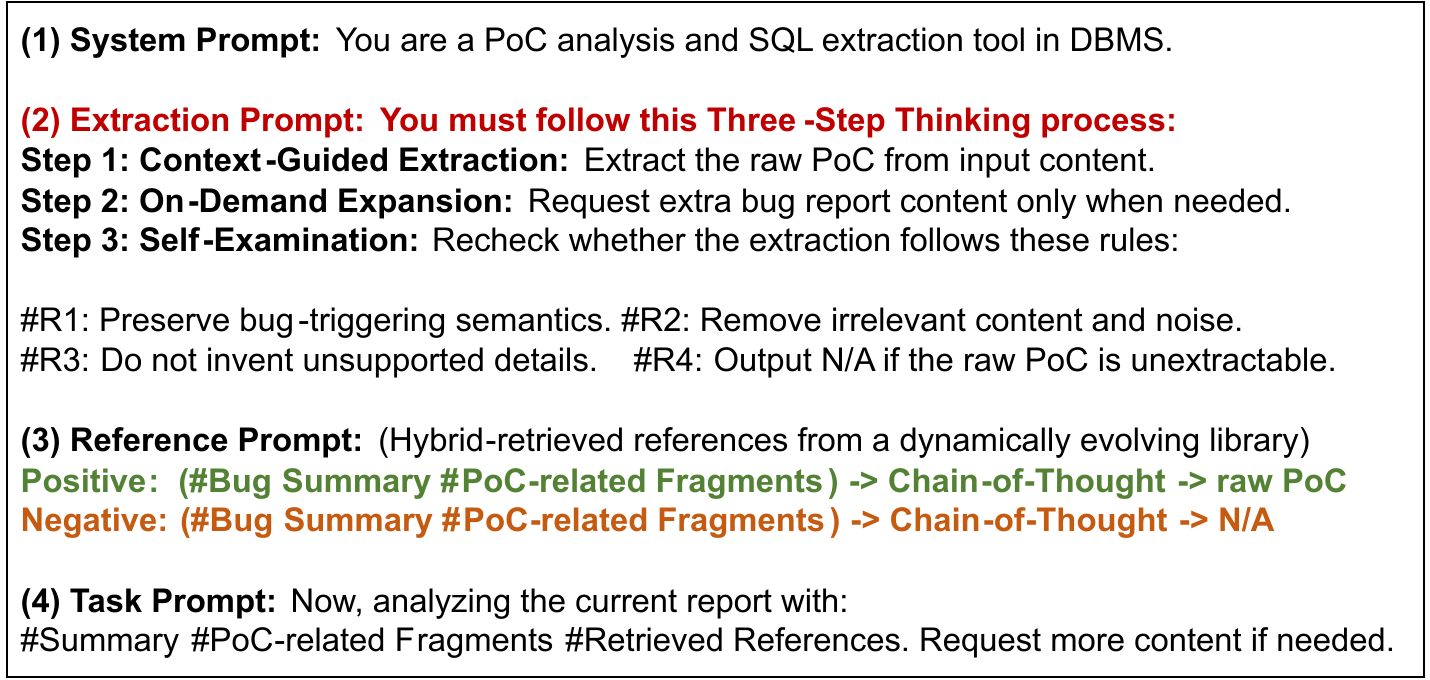}
    \caption{\textbf{An example prompt for LLM in raw PoC extraction.} \mdseries It consists of four components: (1) System Prompt, which instructs LLM to analyze bug reports. (2) Extraction Prompt, which guides the LLM through extraction, on-demand context expansion, and self-examination. (3) Reference Prompt, which retrieves positive and negative exemplars from the library. (4) Task Prompt, which supplies the bug summary, raw PoC-related fragments, and retrieved references for the target report.}
    \label{fig:prompt}
\end{figure}

As illustrated in Figure~\ref{fig:prompt}, \BugForge organizes raw PoC extraction through a structured prompt with four components. The system prompt defines the model as a DBMS-oriented PoC analysis and SQL extraction tool. The extraction prompt specifies a three-step process: context-guided extraction from the bug summary and PoC-related fragments identified in preprocessing, on-demand expansion when the current input is insufficient, and self-examination to verify extraction quality. The reference prompt provides hybrid-retrieved examples, including positive cases with reasoning traces and a negative case for non-extractable reports. Finally, the task prompt instantiates the extraction with the current report summary, PoC-related fragments, and retrieved references.

First, \BugForge builds a compact context from the DBMS bug report summary and the initial PoC-related fragments, and then retrieves relevant exemplars from an automatically maintained reference library. This library is initially seeded with manually curated canonical cases and continuously enriched with newly validated extraction exemplars, enabling \BugForge to maintain a reusable and progressively improving knowledge base for subsequent extraction tasks. As shown in Figure~\ref{fig:prompt}, these retrieved positive and negative exemplars are incorporated into the prompt as reference knowledge. 


To retrieve informative exemplars, \BugForge adopts a hybrid strategy that combines dense retrieval with keyword-based search. 
The dense retriever captures semantic similarity across bug reports, while the keyword retriever emphasizes DBMS-specific lexical cues, such as SQL operators, error messages, and execution symptoms. 
By jointly leveraging semantic relevance and lexical overlap, \BugForge retrieves exemplars similar in both bug semantics and SQL structure. 
As shown in Figure~\ref{fig:prompt}, these exemplars are incorporated as reference context, each including Chain-of-Thought reasoning traces that map input content to raw PoC outputs, providing grounded guidance for extraction.



Next, \BugForge introduces additional report content only when the available evidence is insufficient to determine whether the report contains an extractable raw PoC, or to validate the extracted result using nearby descriptions, execution traces, or stack information. Since the PoC-related fragments may still contain local noise or boundary ambiguity, this adaptive context expansion enables \BugForge to incorporate richer evidence only when necessary, thereby improving extraction precision while avoiding unnecessary token consumption.

\begin{figure}[b]
    \centering\includegraphics[width=0.93\linewidth]{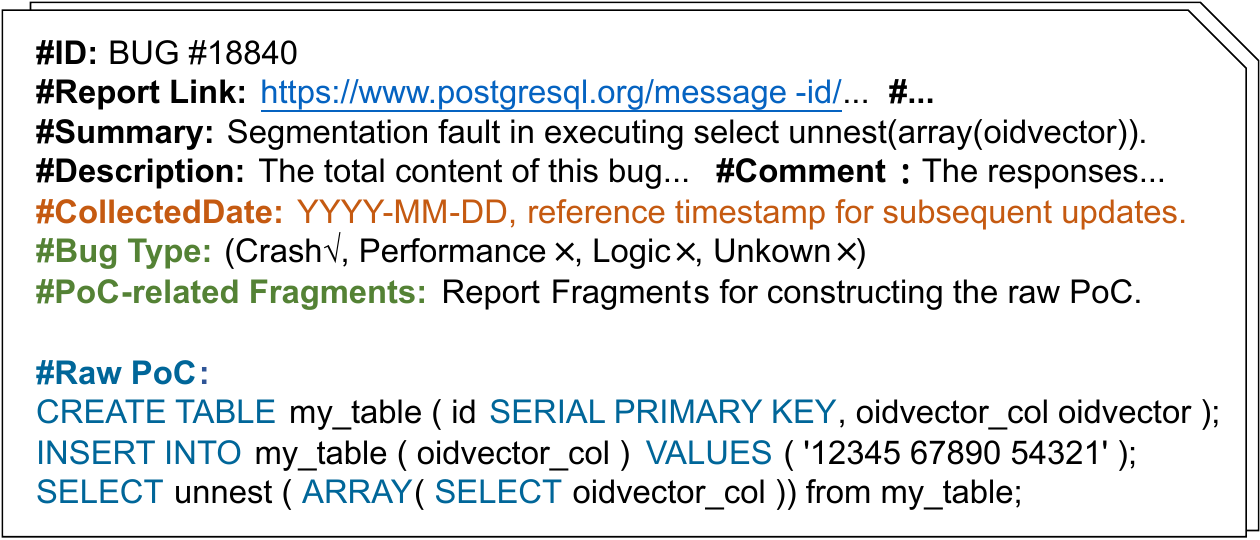}
    \caption{\textbf{An example case in \BugForge repository.} \mdseries
    \BugForge{} normalizes heterogeneous DBMS bug reports into a structured representation. Each case includes: (1) basic metadata of the report, (2) summarized bug-type labels and the PoC-related fragments, (3) the extracted raw PoC used for downstream adaptation and DBMS testing.}
    \label{fig:case}
\end{figure}

Finally, after extracting a raw PoC, \BugForge evaluates its quality in terms of contextual consistency and structural completeness. Specifically, it checks whether the extracted PoC remains consistent with the surrounding report context, whether it preserves the semantics of raw PoCs, and whether it introduces irrelevant SQL. High-confidence cases are then distilled into reusable exemplars and inserted back into the RAG library, allowing the knowledge base to evolve over time and provide increasingly relevant support for subsequent reports; otherwise, the report is classified as non-extractable. The validated raw PoC is then integrated with report metadata to construct the DBMS bug repository, as shown in Figure~\ref{fig:case}.
\subsection{Bug Repository Utilization}

Bug Repository utilization aims to convert raw PoCs into high-quality test cases and then leverages them for automated DBMS testing. Although the repository has been constructed, raw PoCs often suffer from syntactic or semantic errors, as well as missing configuration, which prevent their utilization.
The challenge is to transform them into high-quality test cases by repairing configuration errors, correcting code mistakes, and isolating execution environments to avoid conflicts in different adaptation batches. 

\subsubsection{Semantic-Guided Test Case Adaptation}

Although \BugForge leverages LLMs to extract raw PoCs from bug reports, these PoCs are not always directly utilizable for DBMS testing. 
On the one hand, certain bug reports date back too far, causing the syntax in raw PoCs to be outdated, while others may contain semantic inconsistencies or omit critical configuration information required for execution.
On the other hand, excessive modifications driven solely by the pursuit of executability may reduce the semantic richness compared to raw PoCs. In addition, the execution of one adapted test case may interfere with subsequent ones by altering the database state or leaving residual side effects in the runtime environment.

\begin{figure}[h]
    \centering\includegraphics[width=1\linewidth]{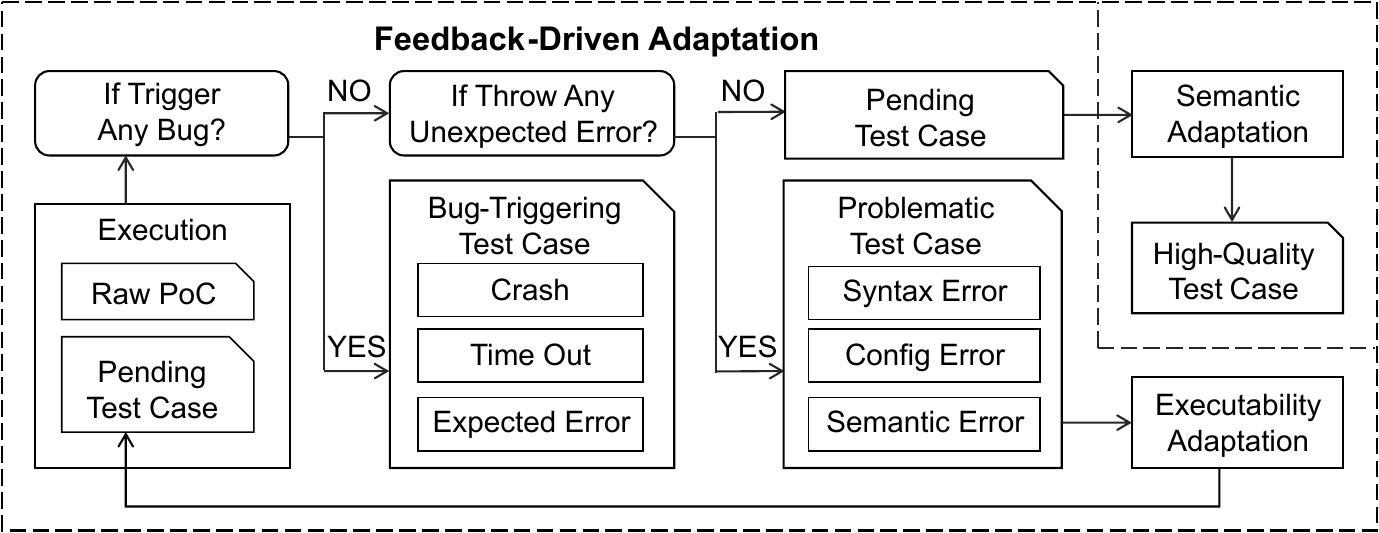}
    \caption{\textbf{Process of Feedback-Driven Adaptation.} \mdseries Each raw PoC is first executed and classified by its observed behavior. Cases with unexpected errors are further diagnosed using execution feedback and DBMS knowledge, and then repaired through LLM-guided executability adaptation.}
    \label{fig:iteration}
\end{figure}

\begin{figure}[h]
    \vspace{-0.2cm}
    \centering\includegraphics[width=0.95\linewidth]{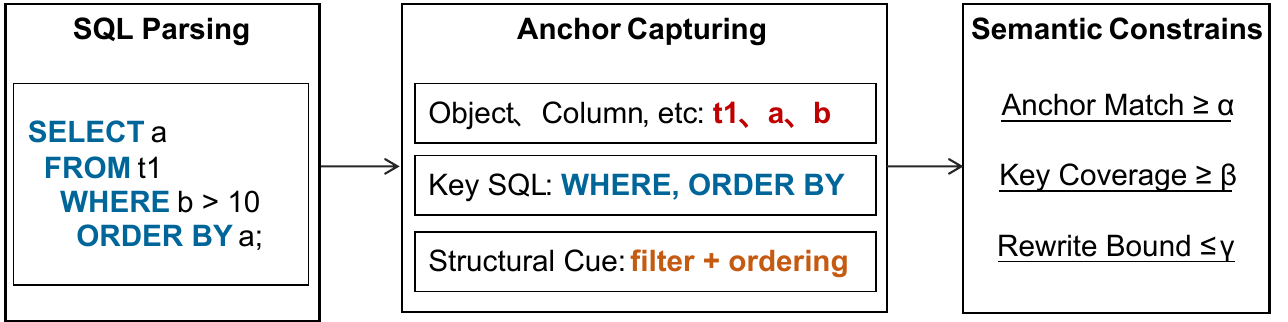}
    \caption{\textbf{Process of Semantic-Constrained Adaptation.} \mdseries \BugForge first extracts semantic anchors from raw PoC, including data dependencies, SQL keywords, and structural cues, and organizes them into an anchor profile. Based on it, \BugForge applies three constraints, namely anchor match, key coverage, and rewrite bound, to guide the LLM in preserving as much semantic richness in raw PoCs as possible during adaptation.}
    \label{fig:check}
\end{figure}

To construct high-quality test cases from raw PoCs, \BugForge adopts a \textit{semantic-guided adaptation} strategy with two levels. It first performs \textit{feedback-driven adaptation }(Figure~\ref{fig:iteration}), which diagnoses execution-blocking issues through runtime feedback and repairs these. Next, \BugForge applies \textit{semantic-constrained adaptation }(Figure~\ref{fig:check}), which uses semantic anchors captured from raw PoCs to guide adaptation towards results that do not lose the semantic richness of raw PoCs.


\textbf{Feedback-Driven Adaptation.}
To improve the executability of the raw PoCs, \BugForge performs feedback-driven adaptation. As shown in Figure~\ref{fig:iteration}, each raw PoC is first executed and examined according to its observed behavior from the target DBMS. Raw PoCs that already trigger expected bug-relevant outcomes, such as crashes, timeouts, or expected errors in report description, are directly accepted as bug-triggering (i.e., high-quality) test cases. Otherwise, \BugForge further checks whether the execution throws any unexpected errors. Test cases that neither trigger the target bug nor raise other errors are marked as pending cases for subsequent processing, while those with unexpected errors are treated as problematic and forwarded to executability adaptation.


For problematic test cases, \BugForge performs failure diagnosis based on execution outputs and auxiliary DBMS knowledge. Specifically, we build a lightweight knowledge base from official DBMS documentation and its source code, and use it to categorize common failures into syntax errors, configuration errors, and semantic errors. Syntax errors indicate malformed SQL statements, configuration errors reflect missing or incompatible runtime settings, and semantic errors correspond to unresolved schema objects, invalid references, or inconsistent query semantics. If an execution message cannot be mapped to any predefined category, the original runtime feedback is preserved as diagnostic context. The resulting feedback is then provided to the LLM as executable-oriented guidance, enabling it to complete missing information and repair execution-blocking issues.

\textbf{Semantic-Constrained Adaptation.}
While feedback-driven adaptation improves executability, overcorrection by the LLM may unintentionally alter the key semantics of the raw PoC. Therefore, \BugForge introduces a semantic-constrained adaptation mechanism, as illustrated in Figure~\ref{fig:check}, to guide the adaptation process with semantic anchors derived from the original raw PoC.

The \emph{semantic anchor} refers to a bug-relevant SQL element or structural cue captured from the raw PoC that reflects its essential semantics. 
As shown in Figure~\ref{fig:check}, \BugForge captures multiple classes of anchors, including data dependencies (e.g., objects, columns, and other schema-level references), SQL keywords and clause-level operators (e.g. \texttt{JOIN}, \texttt{WHERE}, \texttt{ORDER BY}), and the query patterns (e.g. filtering, ordering, or aggregation). 
These anchors provide explicit semantic guidance for adaptation, steering the pending test case, preserving the semantic richness of the raw PoCs rather than a merely executable but irrelevant variant.

Based on the anchor profile, \BugForge performs semantic adaptation under three complementary criteria. Specifically, \emph{Anchor match} is used to constrain the LLM to keep the core data dependencies of the original raw PoCs. \emph{Key coverage} aims to restrict LLM from modifying key SQL operations. \emph{Rewrite bound} is designed to prevent the LLM from excessively rewriting the original query structure. Through these constraints, \BugForge guides the LLM to improve executability while minimizing semantic deviation from the original raw PoCs. The resulting executable test cases through semantic constraints are utilized for downstream DBMS testing.

\textbf{Environment Stabilization.}
During the adaptation process, the SQL statements in different PoCs may interfere with each other. 
For example, executing \texttt{SET GLOBAL innodb\_purge\_stop\_now=ON} in MySQL may prevent the timely recovery of disk space, resulting in the continuous growth of data files.
To prevent this issue, a simple solution is to reinstall the database after the completion of each input case.
However, this approach exhibits poor scalability when executing a large number of cases, as reinstalling the database is time-consuming and unsuitable for actual production.

To strictly guarantee execution independence while maximizing throughput, BugForge employs a statement-driven hierarchical impact assessment mechanism. Prior to execution, this active assessment parses the test inputs into atomic SQL statements and identifies key semantic components, such as operation verbs and target objects, for risk mapping. This mapping drives a differentiated strategy that categorizes operations based on their environmental scope, distinguishing between global-state mutations, system-level configurations, and localized manipulations. Based on this pre-execution risk classification, BugForge dynamically schedules the optimal isolation strategy to be implemented immediately subsequent to a case execution. For instance, if a high-risk pattern (e.g., \texttt{CREATE ROLE} or \texttt{CREATE EXTENSION}) is detected during the initial scan, the system triggers a mandatory container reinstallation post-completion to remediate the altered global state. In contrast, for medium-risk commands like \texttt{ALTER SYSTEM}, the scanner prescribes a state verification or service restart, while low-risk localized manipulations only necessitate lightweight database cleaning.

\vspace{5pt}
\subsubsection{Automated DBMS Testing Enhancement}
The bug repository constructed through \BugForge can be leveraged in various scenarios, among which one of the most significant applications lies in the testing of DBMSs.
In general, DBMSs are equipped with built-in test cases. However, these cases are often relatively simple and thus insufficient for thoroughly exercising the deeper, more complex logic of DBMSs, where bugs are more likely to reside.
To enable comprehensive exploration of DBMSs, \BugForge proposes using raw PoCs from bug reports as the foundation for testing.
We demonstrate how the bug repository constructed by \BugForge can be leveraged in representative DBMS testing workflows, including fuzzing, regression testing, and cross-DBMS testing.

\textbf{DBMS Fuzzing.} 
DBMS fuzzing generates a large number of normal and abnormal test cases for the target DBMS and observes the system’s outputs and state to determine whether bugs are present.
Mutation-based test case generation is a widely used approach in DBMS fuzzing, where existing test cases serve as seeds that are mutated according to syntax rules to produce new test cases. The success of fuzzing is closely tied to seed quality, as higher-quality seeds facilitate faster and more extensive bug discovery.
Employing test cases from \BugForge repository as fuzzing seeds presents two key advantages compared to the built-in test cases provided by DBMS.
Firstly, these test cases have a history of triggering bugs, and areas of code surrounding previously buggy regions are more likely to harbor new bugs. 
Secondly, their relative structural complexity allows the DBMS fuzzer to navigate intricate execution paths to reach deeper and less frequently exercised parts of the code.

\textbf{Regression Testing.}
Regression testing is a fundamental software testing process designed to ensure that new code modifications do not adversely affect existing functionalities or reintroduce previously fixed bugs. 
The bug-triggering test cases can be utilized for regression testing. 
Specifically, by applying these test cases to their corresponding fixed versions and other subsequent long-term release versions, it can be further verified whether historical defects have been thoroughly resolved. 
This approach is better than simply confirming the immediate effectiveness of a repair patch, which can also detect issues such as patch ineffectiveness or bug recurrence caused by version iterations.

\textbf{Cross-DBMS Testing.}
Generally, DBMSs implement the basics of ANSI SQL in their common applications and support more advanced functions through their SQL dialects.
Since different databases within the same series often possess similar foundational architectures and design principles, they are likely to share common bugs. 
In this regard, test cases can provide valuable insights, which can be used to identify potential common bugs that may exist between different DBMSs.
For example, we can test whether bug-triggering cases for MySQL also trigger bugs in MariaDB.

\section{Evaluation}
To evaluate the effectiveness of \BugForge in real-world DBMS testing, we investigate the following research questions to build a DBMS bug repository with \BugForge.

\begin{itemize}
    \item \textbf{RQ1:} What are the characteristics and utilizability of the bug repository constructed by \BugForge?
    \item \textbf{RQ2:} Can \BugForge help discover new DBMS bugs?
    \item \textbf{RQ3:} How does \BugForge{} compare with the state-of-the-art DBMS testing tools?
    \item \textbf{RQ4:} How do high-quality test cases in \BugForge{} compare with others in terms of DBMS testing?
    %
\end{itemize}

\subsection{Implementation and Evaluation Setup}
\textbf{Implementation.} As shown in Figure~\ref{whole_design}, \BugForge consists of two modules: repository construction and repository utilization. The current prototype comprises approximately 17K lines of Python code and 2.8K lines of Shell code. The artifact of \BugForge is available at the anonymous repository: https://anonymous.4open.science/r/BugForge-Artifact-DA6F. In the repository construction module, we implemented database-specific collectors to acquire bug reports from open-source issue trackers and mailing lists, normalize the collected contents into a unified JSON format, and extract SQL-oriented raw PoCs through a filtering stage. In the repository utilization module, we implemented containerized execution backends and supporting automation scripts for environment setup, health checking, database cleanup, restart/reset, and dry-run validation, together with the components for PoC extraction, adaptation, and seed smelting. For the LLM-assisted stages, we employ Gemini-2.5-flash~\cite{geminiflash} for raw PoC extraction and Gemini-2.5-pro~\cite{geminipro} for the more complex adaptation tasks. 

%






\textbf{Tested DBMSs.} To evaluate the effectiveness and applicability of \BugForge{}, we select four representative open-source DBMSs as evaluation targets: \mysql{}~\cite{mysql}, \pg{}~\cite{pg}, \mariadb{}~\cite{mariadb}, and \monetdb{}~\cite{monetdb}.
These systems are widely used in research and industry and span diverse query engines, storage models, and system architectures. 
In addition, these DBMSs have long served as standard subjects in prior DBMS testing research~\cite{sqlancer, apollo, sqlancer-norec,griffin,sedar}.

\textbf{Basic Setup.} The evaluation was conducted on a server running 64-bit Ubuntu 22.04.5 LTS, equipped with an AMD EPYC 7763 64-core processor (128 threads) and 500 GB of main memory. 
For DBMS fuzzing and cross-DBMS testing, the targeted DBMSs were compiled with AFL++~\cite{aflpp_parallel} to enable testing support. We ran the DBMS fuzzers under their default configurations together with their corresponding sets of initial seeds. Each fuzzing instance was executed for 24 hours on the target DBMS.
For regression testing, we selected the latest available version as well as the fixed versions supported by each target DBMS to examine whether historical bugs reappeared in later releases.
To identify similar bugs across systems, we further performed cross-testing between MySQL and MariaDB by applying the test cases of one system to the other for 24 hours. In addition, both fuzzing and cross-DBMS testing were repeated three times.

\subsection{Assessment of \BugForge{} Bug Repository}
\textbf{Characteristic of Bug Repository.} 
Table~\ref{tab:dbms_statistics} summarizes the characteristics of the bug repository constructed by \BugForge{} in several aspects.
For \textit{covered time}, the bug repository encompasses all publicly available bug reports for PostgreSQL, MySQL, MariaDB, and MonetDB spanning from \textbf{\timerange{}}.
For \textit{overall statistics}, the repository contains {\totalreport} bug reports, including \mysqlreport{} from \mysql{}, \mariadbreport{} from \mariadb{}, \pgreport{} from \pg{}, and \monetdbreport{} from \monetdb{}, respectively.
From these reports, \totalrawpoc{} raw PoCs were extracted.
The total number of raw PoCs is slightly lower than the total number of reports, as some bug reports do not include PoC-related text or code snippets.

For \textit{security vulnerability coverage}, the repository covers 2,597 CVE reports, including 2,055 related to \mysql{}, 147 related to MariaDB, 364 related to \pg{}, and 31 related to \monetdb{}. 
The CVSS-based severity distribution of these CVEs spans four severity levels, including 158 critical, 502 high, 1,681 medium, and 256 low severity vulnerabilities.
Based on these raw PoCs, \BugForge{}  provides \mysqlexecutabletestcases, \mariadbexecutabletestcases, \pgexecutabletestcases and \monetdbexecutabletestcases high-quality test cases for MySQL, MariaDB, PostgreSQL, and MonetDB, respectively.





\begin{table}[htbp]
    \centering
    \caption{Statistics of Bug Repository Built by \BugForge{}}
    \label{tab:dbms_statistics}
    
    \renewcommand{\arraystretch}{1.2} 
    \setlength{\aboverulesep}{0pt}
    \setlength{\belowrulesep}{0pt}
    \setlength{\tabcolsep}{2pt}
    
    \resizebox{1\linewidth}{!}{ 
    \begin{tabular}{l | c c c c | c}
        \toprule
        \textbf{} & \textbf{MySQL} & \textbf{MariaDB} & \textbf{PostgreSQL} & \textbf{MonetDB} & \textbf{Total} \\
        \midrule
        Covered Time & 2003-2026 & 2009-2026 & 1998-2026 & 2011-2026 & -- \\
        Collected Reports & 18,212 & 12,442 & 4,163 & 2,815 & 37,632 \\
        Extracted Raw PoCs & 17,723 & 11,581 & 3,758 & 2,468 & 35,530 \\
        CVEs & 2,055 & 147 & 364 & 31 & 2,597\\
        \midrule
        high-quality Test Cases & 16,112 & 11,426 & 3,616 & 2,444 & 33,598 \\
        \bottomrule
    \end{tabular}
    }
\end{table}

\textbf{Utilizability of Bug Repository.} 
We assess the usability of the bug repository by evaluating the coverage of SQL features and the number of adapted PoCs it contains.

\textit{SQL Feature Coverage.} Table~\ref{tab:system_coverage_statistics} quantifies the comprehensiveness of the repository by using the total count of data types and SQL keywords extracted from internal DBMS system tables as a baseline. 
The statistics reveal that the test cases in \BugForge achieve a coverage rate of 88.62\% for data types and 89.82\% for keywords across all DBMSs.
Notably, the executable test cases cover all of the built-in data types for MySQL and over 90 percent of the keywords for both MariaDB and PostgreSQL. 
These results demonstrate that the cases in \BugForge effectively encapsulate a broad spectrum of core language constructs and system-specific features.

\begin{table}[htbp]
    \centering
    \caption{Covered Data Types and Keywords in DBMSs}
    \label{tab:system_coverage_statistics}
    \setlength{\aboverulesep}{0.5pt}
    \setlength{\belowrulesep}{0.5pt}
    \setlength{\tabcolsep}{4pt}
    \resizebox{\linewidth}{!}{
    \begin{tabular}{l | c c c | c c c}
        \toprule
        \multirow{2}{*}{\textbf{DBMS}} & \multicolumn{3}{c|}{\textbf{Data Type}} & \multicolumn{3}{c}{\textbf{Keyword}} \\
        \cmidrule(lr){2-4} \cmidrule(lr){5-7}
        & Total & Covered & Rate & Total & Covered & Rate \\
        \midrule
        MySQL      & 33  & 33  & 100.00\%       & 754   & 666   & 88.33\% \\
        MariaDB    & 36  & 35  & 97.22\%        & 683   & 616   & 90.19\% \\
        MonetDB    & 30  & 28  & 93.33\%        & 301   & 268   & 89.04\% \\
        PostgreSQL & 68  & 52  & 76.47\%        & 491   & 452   & 92.06\% \\
        \midrule
        \textbf{Total} & \textbf{167} & \textbf{148} & \textbf{88.62\%} & \textbf{2229} & \textbf{2002} & \textbf{89.82\%} \\
        \bottomrule
    \end{tabular}
    }
\end{table}

\textit{high-quality Test Cases Generated by \BugForge{}.}  
Among the raw PoCs, 16,300 cases can be directly used as high-quality test cases, the remaining 19,230 cases are treated as problematic during adaptation.
Table~\ref{tab:repair_performance_final} presents the number of high-quality test cases generated through adaptation for DBMS testing.
Here, high-quality test cases refer to problematic raw PoCs that successfully pass both feedback-driven adaptation and semantic-constrained adaptation, meaning that they are executable and satisfy the semantic anchor constraints.
Under this criterion, the approach converted a total of 17,298 cases into high-quality test cases, accounting for 89.95\% of the total.
\begin{table}[htbp]
    \centering
    \caption{Number of High-quality Test Cases Adapted from Problematic Test Cases.}
    \label{tab:repair_performance_final}
    \renewcommand{\arraystretch}{1.1}
    \setlength{\tabcolsep}{2pt}
    \resizebox{1\linewidth}{!}{
    \begin{tabular}{l | c c c c | c}
        \toprule
        & \textbf{MySQL} & \textbf{MariaDB} & \textbf{PostgreSQL} & \textbf{MonetDB} & \textbf{Total} \\
        \midrule
        Problematic Cases & 12,504 & 4,523 & 1,911 & 292 & \textbf{19,230} \\
        Adaptable Cases & 10,893 & 4,368 & 1,769 & 268 & \textbf{17,298} \\
        \midrule
        Ratio & 87.12\% & 96.57\% & 92.57\% & 91.78\% & \textbf{89.95\%} \\
        \bottomrule
    \end{tabular}
    }

\end{table}
\vspace{-0.5cm}
\subsection{Bugs Detected With \BugForge}



\newcolumntype{L}{>{\RaggedRight\arraybackslash}X}


\begin{table*}[t!]

\caption{List of previously unknown bugs detected by DBMS testing with \BugForge{} employed}
\label{tab:bug_summary_full}

\centering
\footnotesize 

\setlength{\tabcolsep}{3.2pt} 

\renewcommand{\arraystretch}{1}
\setlength{\aboverulesep}{1pt}
\setlength{\belowrulesep}{1pt}

\begin{tabularx}{\linewidth}{c c c c c c X}
    \toprule
    \textbf{\#} & \textbf{DBMS} & \textbf{ID} & \textbf{Status} & \textbf{Component} & \textbf{Testing Method} & \textbf{Description}\\
    \midrule

    1 & \mysql & S2310126 & Confirmed & Function & DBMS Fuzzing & MySQL 9.4.0 crashes at Item\_func\_bit::val\_int.\\
    2 & \mysql & S2310135 & Confirmed & Function & DBMS Fuzzing & MySQL 9.4.0 crashes at Item\_func::print\_op.\\
    3 & \mysql & S2310142 & Confirmed & Function & DBMS Fuzzing & MySQL 9.4.0 crashes at Item\_func::walk.\\
    4 & \mysql & S2310157 & Confirmed & Item & DBMS Fuzzing & MySQL 9.4.0 crashes Item\_ref::walk.\\
    5 & \mysql & S2310161 & Confirmed & Item & DBMS Fuzzing & MySQL 9.4.0 crashes at Item\_ref::val\_int.\\
    6 & \mysql & S2310174 & Confirmed & Filed & DBMS Fuzzing & MySQL 9.4.0 crashes at Field::set\_notnull.\\
    7 & \mysql & S2310188 & Confirmed & Field & DBMS Fuzzing & MySQL 9.4.0 crashes at set\_field\_to\_null\_with\_conversions.\\
    8 & \mysql & S2310190 & Confirmed & Optimizer & DBMS Fuzzing & MySQL 9.4.0 crashes at Query\_expression::optimize.\\
    9 & \mysql & S2310208 & Confirmed & View & DBMS Fuzzing & Server crashes at Item\_view\_ref::check\_column\_privileges.\\
    10 & \mysql & \# 119085 & Reported & Server & Cross-DBMS Testing & FLUSH PRIVILEGES crash after table column drop.\\
    11 & \mysql & \# 119086 & Reported & Stored Routines & Cross-DBMS Testing & Server crash after VIEW-to-TABLE dependency change.\\
    12 & MariaDB & MDEV-37625 & Confirmed & Server & DBMS Fuzzing &  MariaDB 12.1.1-rc crashes at alloc\_root.\\
    13 & MariaDB & MDEV-37628 & Confirmed & Optimizer & DBMS Fuzzing &  MariaDB 12.1.1-rc crashes at find\_field\_in\_tables.\\
    14 & MariaDB & MDEV-37640 & Confirmed & JSON/Server & DBMS Fuzzing & MariaDB 12.1.1-rc crashes at String::append.\\
    15 & MariaDB & MDEV-37641 & Confirmed & Optimizer & DBMS Fuzzing & MariaDB 12.1.1-rc crashes at sub\_select\_postjoin\_aggr.\\
    16 & MariaDB & MDEV-37646 & Confirmed & Optimizer & DBMS Fuzzing & MariaDB 12.1.1-rc crashes at Item::save\_decimal\_in\_field.\\
    17 & MariaDB & MDEV-37647 & Confirmed & JSON/Optimizer/Server & DBMS Fuzzing & Server crashes at set\_field\_to\_null\_with\_conversions.\\
    18 & MariaDB & MDEV-37648 & Confirmed & Optimizer/Server & DBMS Fuzzing & Server crashes at Item\_direct\_view\_ref::val\_decimal.\\
    19 & MariaDB & MDEV-37768 & Reported & Server & Cross-DBMS Testing & Server crash when mysql.procs\_priv has an invalid structure.\\
    20 & MariaDB & MDEV-37671 & Confirmed & 
Optimizer/Server & Regression Testing & Solved bug MDEV-32326 reappeared in 12.1.1-rc.\\
    21 & MariaDB & MDEV-37735 & Confirmed & Sequences & Regression Testing & Solved bug MDEV-37172 reappeared in 11.8.4 and 12.1.1-rc.\\
    22 & MariaDB & MDEV-37734 & Reported & Optimizer & Regression Testing & Solved bug MDEV-32308 reappeared in 11.8.4 and 12.1.1-rc.\\
    23 & MonetDB & \# 7720 & Confirmed & Server & DBMS Fuzzing & MonetDB Mar2025-SP2 server crashes at stmt\_cond.\\
    24 & MonetDB & \# 7721 & Reported & Server & DBMS Fuzzing & MonetDB Mar2025-SP2 server crashes at stmt\_replace.\\
    25 & MonetDB & \# 7722 & Fixed & Server & DBMS Fuzzing & MonetDB Mar2025-SP2 server crashes at rel\_with\_query.\\
    26 & MonetDB & \# 7723 & Reported & Server & DBMS Fuzzing & MonetDB Mar2025-SP2 server crashes at rel\_has\_freevar.\\
    27 & MonetDB & \# 7724 & Reported & Server & Regression Testing & Solved bug \# 3009 reappeared in Mar2025-SP2 release.\\
    28 & MonetDB & \# 7725 & Fixed & Server & Regression Testing & Solved bug \# 7482 reappeared in Mar2025-SP2 release.\\
    29 & MonetDB & \# 7726 & Reported & Server & Regression Testing & Solved bug \# 7486 reappeared in Mar2025-SP2 release.\\
    30 & MonetDB & \# 7727 & Fixed & Server & Regression Testing & Solved bug \# 7436 reappeared in Mar2025-SP2 release.\\
    31 & MonetDB & \# 7728 & Reported & Server & Regression Testing & Solved bug \# 7488 reappeared in Mar2025-SP2 release.\\
    32 & MonetDB & \# 7729 & Confirmed & Server & Regression Testing & Solved bug \# 7472 reappeared in Mar2025-SP2 release.\\
    33 & PostgreSQL & \# 19055 & Confirmed & Parser & DBMS Fuzzing & The count() aggregate is assigned the
wrong agglevelsup.\\
    34 & PostgreSQL & \# 19382 & Fixed & Executor & DBMS Fuzzing & PostgreSQL 14-17.7 server crash at \_\_nss\_database\_lookup.\\
    35 & PostgreSQL & \# 19384 & Reported & Parser & DBMS Fuzzing & PostgreSQL 17.7 server crashes at textout.\\

\bottomrule
\end{tabularx}

\end{table*}


The four tested DBMSs are widely used in industry and well tested by existing DBMS testing tools, making it difficult to find new bugs. Nevertheless, our evaluation shows that historical bugs captured by \BugForge{} can still be leveraged to uncover new or insufficiently addressed bugs through three types of testing: DBMS fuzzing, DBMS regression testing, and cross-DBMS verification. 

\textbf{Testing Scenarios.}  For \textit{DBMS fuzzing}, we reuse the high-quality test cases generated by \BugForge as initial seeds to integrate into the pipelines of \griffin{}~\cite{griffin}. 
For \textit{DBMS regression testing}, we replay bug-triggering test cases on newer versions in \mariadb{} and \monetdb{} to identify whether past bugs have been properly fixed.
A regression bug is counted when a test case still triggers the same issue in versions marked as fixed.
For \textit{cross-DBMS verification}, we leverage the high syntactic and architectural similarity between \mysql{} and \mariadb{}.
Test cases produced by \BugForge{} for one DBMS are used as initial seeds for \griffin{} to test the other.


\textbf{Detected Bugs.} Table~\ref{tab:bug_summary_full} summarizes the number of detected bugs in the four DBMS: \pgbug in \pg{}, \mysqlbug in \mysql{}, \mariadbbug in \mariadb{}, and \monetdbbug in \monetdb{}. Among the detected \totalBugs bugs by \BugForge, 23 are detected through DBMS fuzzing, 9 through regression testing, and 3 through cross-DBMS testing. 
We have responsibly disclosed all newly discovered bugs to the corresponding DBMS vendors. At the time of writing this paper, \totalConfirmedBugs bugs have been confirmed. Among the confirmed bugs, 9 are reserved by vendors for internal patching. 
Some bugs have received positive feedback from developers.
One developer commented in detected \pg bug \#19382 \cite{pg19382},``\textit{In fact, this fails as fast as 'SELECT bar()' for me. Also, it fails for REL\_14\_STABLE, REL\_16\_STABLE, so, problem is for all supported versions}''.

Among the detected bugs, 22 are classified as high-risk vulnerabilities by the DBMS developers.
The identified issues are distributed across a variety of critical components, revealing bugs at multiple layers of the database architecture. 
For instance, bugs in MySQL and MariaDB affect core areas such as the Optimizer, JSON/Server, and low-level data functions, while the PostgreSQL bug was pinpointed in its Parser.
Several issues can be triggered by minimal inputs and reflect persistent blind spots in error-handling routines. 
Furthermore, the analysis highlights persistent challenges in regression testing, as several previously solved bugs were observed to have reappeared in subsequent releases of both \mariadb{} and \monetdb{}.
These results confirm that reusing real-world bug knowledge from \BugForge{} not only enhances DBMS testing effectiveness but also helps uncover deep-rooted defects across versions and systems.

\begin{figure}[h]
    \centering\includegraphics[width=0.8\linewidth]{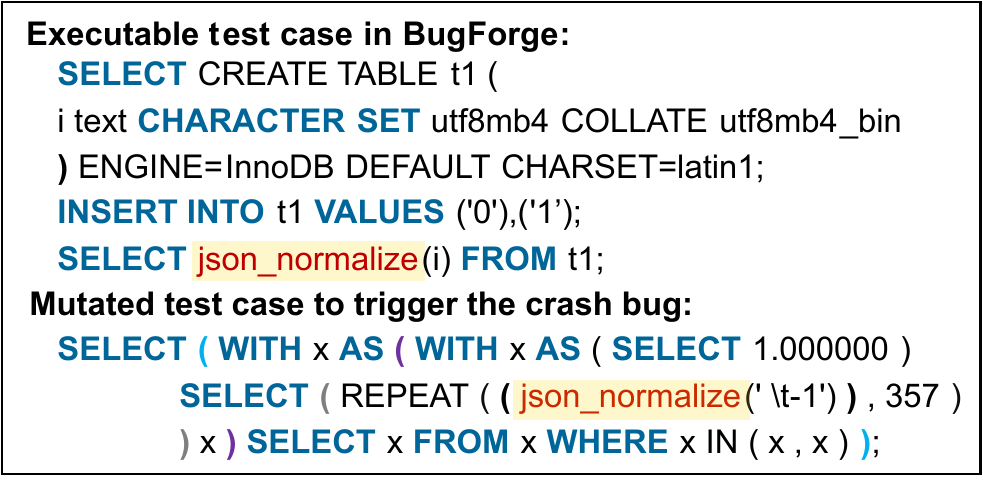}
    \caption{\textbf{A crash bug in MariaDB found by fuzzing.}}
    \label{fig:case_1}
    \vspace{-0.2cm}
\end{figure}

\textbf{Case Study 1: \BugForge{} for DBMS Fuzzing.} As shown in the lower part of Figure~\ref{fig:case_1}, \BugForge detected a buffer overflow bug in MariaDB, triggered by a \texttt{SELECT} query with the \texttt{REPEAT} function. Ordinarily, the REPEAT function correctly allocates memory when its input is a constant string. However, in our case, the optimizer cannot accurately predict the output size of \texttt{json\_normalize}, leading it to allocate an insufficiently sized buffer for the \texttt{REPEAT} function.
This size mismatch results in a buffer overflow during the string repetition process, ultimately causing the server to crash.

Notably, among the \mariadbreport{} reports collected from \mariadb{}, only 4 raw PoCs involve the function \texttt{json\_normalize}.
For example, the historical report MDEV-29026 describes a use-after-poison issue in \texttt{json\_normalize\_number}, providing insight into the potential instability of \texttt{json\_normalize} in handling memory and data types.
In our experiment, mutator reallocates pressure from the internal parsing logic of function \texttt{json\_normalize} to the interactions among different server components, thereby exploring the deep state space of MariaDB and uncovering new bugs.
In other words, it would be quite difficult for DBMS fuzzers to discover such a bug without this kind of case as input guidance.
This case demonstrates the efficacy of leveraging historical bug data to detect new bugs.

\begin{figure}[h]
    \centering\includegraphics[width=0.9\linewidth]{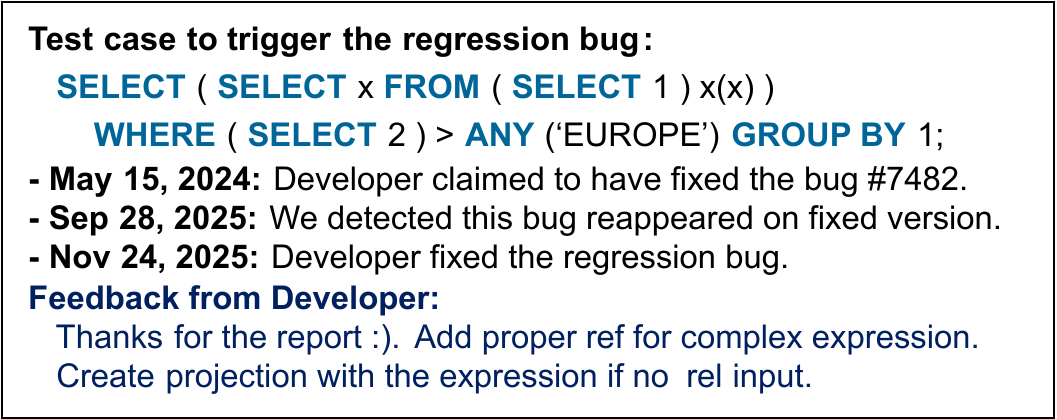}
    \caption{A crash bug in MonetDB found by regression testing.}
    \label{fig:case_2}
    \vspace{-0.25cm}
\end{figure}

\textbf{Case Study 2: \BugForge{} for Regression Testing.} Figure~\ref{fig:case_2} shows a confirmed regression bug in \monetdb{} detected by \BugForge{}. The bug manifests as a system failure triggered by a nested query that involves a specific combination of \texttt{GROUP BY} and the \texttt{ANY} comparison operator applied to a string literal.

Although this bug appeared in the recent "Mar2025-SP2" release, where such historical execution logic was expected to be stable, \BugForge{} successfully reproduced the failure using the executable test case.
Upon receiving our report, MonetDB developers confirmed the regression and promptly issued a fix (referencing issue \#7725).
The root cause analysis revealed a defect in the query projection phase: the optimizer failed to add proper references for complex expressions when no relational input was present.
This case underscores \BugForge{}'s capability to enforce rigorous regression testing, ensuring that patches for complex logic errors are not only applied but also maintained correctly across subsequent release branches.

\begin{figure}[h]
    \centering\includegraphics[width=0.9\linewidth]{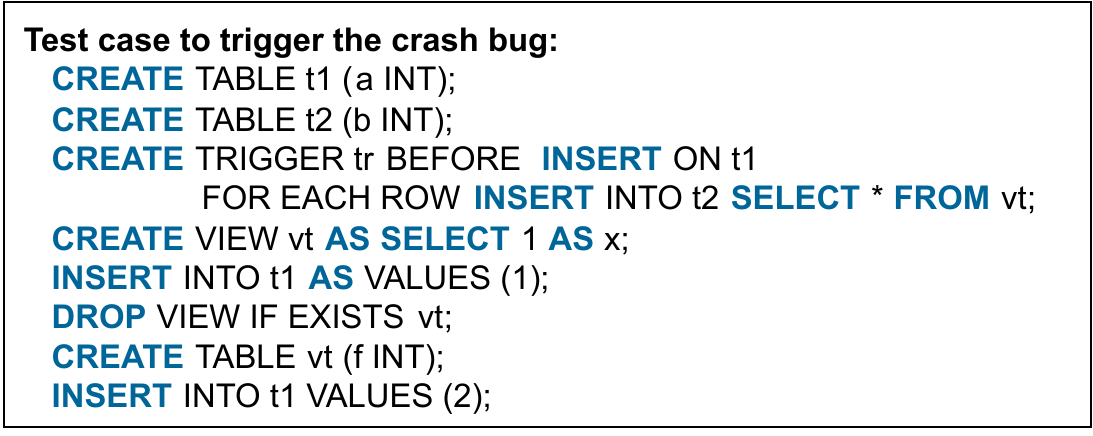}
    \caption{A crash bug in MySQL found by cross-DBMS testing.}
    \label{fig:case_3}
    \vspace{-0.25cm}
\end{figure}

\textbf{Case Study 3: \BugForge{} for Cross-DBMS Testing.} Figure~\ref{fig:case_3} details a representative confirmed bug found during the MariaDB2MySQL campaign to highlight the practical value of our approach. 
The bug causes a server crash in the latest MySQL versions through a specific interaction between Triggers and Views: 
The crash occurs when a View referenced by a Trigger is dropped and recreated as a Table, exposing a flaw in MySQL’s handling of metadata dependencies for stored routines during schema changes.

This discovery emphasizes the power of syntactic compatibility. A test case originally designed to verify logical correctness in MariaDB was able to traverse the valid SQL parser in MySQL and trigger a fatal error in the underlying execution engine. 
This confirms that utilizing seeds from a sibling DBMS allows automated testing to explore complex state transitions, such as type confusion in dependencies that are rarely covered by isolated fuzzing tools.
This case underscores \BugForge’s ability to detect hidden bugs through cross-DBMS testing, which could detect hidden bugs within bug PoCs from other DBMS.

\subsection{Comparison of State-of-the-art DBMS Testing Tools.}
To further evaluate the practical effectiveness of \BugForge{}, we compare it with two representative DBMS testing tools, \emph{SQLancer} and \emph{SQLsmith}. For \BugForge{}, we use the configuration where Griffin is seeded with both built-in unit test cases and the high-quality test cases generated from \BugForge{} (i.e., \griffinUB{}). For SQLancer, we enable its FUZZ mode for testing. For SQLsmith, we use its default version to test MonetDB and PostgreSQL, and adopt its MySQL branch to test MySQL and MariaDB. Since SQLancer does not support MonetDB, the total is computed only over the three supported DBMSs, i.e., MySQL, MariaDB, and PostgreSQL.

\begin{table}[h]
\centering
\caption{Comparison of Typical DBMS Testing Tools.}
\label{tab:coverage_bug_compare}
\resizebox{\linewidth}{!}{
\begin{tabular}{l|ccc|ccc}
\toprule
\multirow{2}{*}{\textbf{DBMS}}
& \multicolumn{3}{c|}{\textbf{Covered Branches}}
& \multicolumn{3}{c}{\textbf{Detected Bugs}} \\
& SQLancer & SQLsmith & BugForge & SQLancer & SQLsmith & BugForge \\
\midrule
MySQL      & 106,971 & 96,311 & 209,720 & 5 & 3 & 8 \\
MariaDB    & 46,298  & 30,896 & 140,677 & 4 & 2 & 7 \\
PostgreSQL & 48,030  & 50,285 & 90,763  & 2 & 1 & 3 \\
MonetDB    & N/A     & 33,209 & 83,420  & N/A & 2 & 5 \\
\midrule
\textbf{Total$^\dagger$}    
           & 201,299 & 177,492 & 441,160 & 11 & 6 & 18 \\
\bottomrule
\end{tabular}
}
\vspace{1mm}
\end{table}

Table~\ref{tab:coverage_bug_compare} presents the comparison results. Overall, \BugForge{} outperforms both SQLancer and SQLsmith in branch coverage across all evaluated DBMSs. Specifically, \BugForge{} covers 209,720 branches on MySQL, 140,677 on MariaDB, 90,763 on PostgreSQL, and 83,420 on MonetDB, all of which are substantially higher than the corresponding results of the other two tools. Over the three commonly supported DBMSs, \BugForge{} covers 441,160 branches in total, compared with 201,299 for SQLancer and 177,492 for SQLsmith. These results suggest that the high-quality test cases generated from DBMS bug reports can guide testing toward deeper and broader execution paths than purely generator-based approaches.

\textbf{Detected Bugs.} As shown in Table~\ref{tab:coverage_bug_compare}, \BugForge{} detected 8, 7, 3, and 5 bugs on MySQL, MariaDB, PostgreSQL, and MonetDB, respectively. Over the three jointly supported DBMSs except MonetDB, \BugForge{} detected 18 bugs in total, compared with 11 for SQLancer and 6 for SQLsmith. This advantage is partly attributed to the high-quality test cases provided by \BugForge{}, which were derived from DBMS bug reports and thus retain richer bug-relevant structures and historical failure patterns. Overall, the results suggest that leveraging adapted real-world bug reports can serve as a high-quality complement to existing DBMS testing techniques.

\subsection{Comparison of Test Case Quality}
To evaluate the quality of the test cases adapted by \BugForge{}, we leverage its high-quality cases as initial seeds for \griffin{}~\cite{griffin} and \squirrel{}~\cite{squirrel} fuzzing (labeled as  \griffinB{} and \squirrelB{}), 
and compare the results with 
(1) built-in unit test cases from DBMS repositories, labeled as  \griffinU{} and \squirrelU{} in the evaluation.
(2) the high-quality test cases generated from repositories by \BugForge, combined with built-in cases, labeled as  \griffinUB{} and \squirrelUB{} in the evaluation.
We measured their effectiveness by the number of branches and detected bugs. 
The cells corresponding to \squirrel{} on MonetDB are marked as N/A because MonetDB is currently not supported by \squirrel{}.

\begin{table}[htbp]
\centering
\caption{Covered Branches with different initial seeds. }
\label{tab:branch_coverage_final}
\renewcommand{\arraystretch}{1.15} 
\setlength{\tabcolsep}{0.5pt}

\resizebox{1\linewidth}{!}{ 
\begin{tabular}{l | c c c | c c c}
    \toprule
    \textbf{DBMS} & \textbf{\griffinU{}} & \textbf{\griffinB{}} & \textbf{\griffinUB{}} & \textbf{\squirrelU{}} & \textbf{\squirrelB{}} & \textbf{\squirrelUB{}} \\
    \midrule
    MySQL   & 182,709 & 192,710 & 209,720 & 147,936 & 159,794 & 175,794 \\
    MariaDB & 114,029 & 134,595 & 140,677 & 89,115  & 100,188 & 109,318 \\
    PostgreSQL      & 73,273  & 85,862  & 90,763  & 59,253  & 63,235  & 74,828  \\
    MonetDB & 32,990  & 79,797  & 83,420  & N/A      & N/A      & N/A      \\
    \midrule
    \textbf{Total}     & 403,001 & 492,964 & 524,580 & 296,304 & 323,217 & 359,940 \\
    \midrule
    \textbf{Increment} & --      & +22.32\% & +30.17\% & --      & +9.08\% & +21.48\% \\
    \bottomrule
\end{tabular}
}
\end{table}
\vspace{-0.1cm}

\textbf{Branch Coverage.} Table~\ref{tab:branch_coverage_final} presents the number of covered code branches of DBMS fuzzers on DBMSs when using the unit test cases and test cases by \BugForge as initial seeds. 
We can see that DBMS fuzzers equipped with test cases from \BugForge outperform the original ones in terms of branch coverage.
Specifically, compared to \griffinU{}, \mbox{\griffinB{}} and \mbox{\griffinUB{}} cover 22.32\% and 30.17\% more branches in total, respectively. Similarly, compared to \squirrelU{}, \squirrelB{} and \squirrelUB{} cover 9.08\% and 21.48\% more branches, respectively.

These improvements can be attributed to the high-quality test cases provided by the \BugForge{}. Notably, the \BugForge{} corpus contains only about 14.2~MB of total file content, compared with 100.3~MB for the unit-test corpus, corresponding to an 85.9\% reduction.
Despite using a much smaller corpus, \BugForge{} still improves total branch coverage by 22.32\% for \griffinB{} over \griffinU{} and 9.08\% for \squirrelB{} over \squirrelU{}. 
Compared to the unit test cases used by \squirrel{} and \griffin{}, the test cases curated from \BugForge cover a significantly broader range of SQL syntax features and functional scenarios. This enhanced diversity enables the fuzzers to exercise deeper and wider portions of the DBMS codebase, thereby covering more code branches.


\begin{table}[htbp]
    \centering
    \caption{Detected Bugs with different initial seeds.}
    \label{tab:bug-compare}
    
    \renewcommand{\arraystretch}{1.1} 
    \setlength{\tabcolsep}{0.5pt}
    
    \resizebox{1\linewidth}{!}{ 
    \begin{tabular}{l | c c c | c c c}
        \toprule
        \textbf{DBMS} & \textbf{\griffinU{}} & \textbf{\griffinB{}} & \textbf{\griffinUB{}} & \textbf{\squirrelU{}} & \textbf{\squirrelB{}} & \textbf{\squirrelUB{}} \\
        \midrule
        MySQL   & 4 & 6 & 8 & 3 & 5 & 6 \\
        MariaDB & 3 & 5 & 7 & 2 & 4 & 5 \\
        PostgreSQL & 1  & 2  & 3  & 1  & 1  & 2  \\
        MonetDB & 2  & 5  & 5  & N/A      & N/A      & N/A      \\
        \midrule
        \textbf{Total}     & 10 & 18 & 23 & 6 & 10 & 13 \\
        \midrule
        \textbf{Increment} & --      & +80.00\% & +130.00\% & --      & +66.67\% & +116.67\% \\
        \bottomrule
    \end{tabular}
    }
    \vspace{-0.1cm}
\end{table}

\textbf{Detected Bugs.}
Table \ref{tab:bug-compare} exhibits the bugs detected by DBMS fuzzers on DBMSs when using the unit
test cases and test cases by \BugForge as initial seeds.
From the table, we can see that \BugForge{} improves the bug detection ability of \griffin{} and \squirrel{} with unit test cases.
With \BugForge{} applied, \griffinB{} and \griffinUB{} found a total of 8 and 13 more bugs in the four DBMSs than \griffinU{}, respectively; similarly, \squirrelB{} and \squirrelUB{} found 4 and 7 more bugs than \squirrelU{}, respectively.
The improvement is mainly attributed to the broader coverage provided by the test cases from \BugForge{}.
These test cases, generated from raw PoCs, encapsulate a wide range of historical bug-triggering scenarios and types.
When integrated into \griffin{} and \squirrel{}, the mutation-based testing is guided toward previously vulnerable code paths and execution contexts, thereby increasing the likelihood of rediscovering similar bugs or uncovering new ones in adjacent regions.

\section{Discussion}
\textbf{Impact of Adaptation Strategy.}
In \BugForge{}, problematic raw PoCs need to be transformed into high-quality test cases, requiring the adaptation process to improve executability while preserving semantic richness. To achieve this goal, \BugForge{} combines feedback-driven adaptation with semantic-constrained adaptation.
To better understand the effect of different strategies, we compare three adaptation strategies on the same test sets: \emph{F}, which refers to feedback-driven adaptation; \emph{S}, which refers to semantic-constrained adaptation; and their combination, \emph{F + S}, i.e., the strategy adopted by \BugForge{}. For executability and semantic richness, we randomly sampled 1,000 problematic cases, consisting of 250 cases from each of MySQL, MariaDB, PostgreSQL, and MonetDB. For semantic consistency, we further randomly sampled 100 executable cases, consisting of 25 cases from each DBMS, and manually inspected them. The semantic richness metric indicates whether an adapted case can satisfy the semantic-anchor constraints, reflecting that it still preserves decent semantic similarity to the corresponding raw PoC, while semantic consistency is determined by manual inspection.
\begin{table}[htbp]
\centering
\caption{Impact of Different Adaptation Strategies.}
\label{tab:stage_strategy}
\resizebox{\linewidth}{!}{
\begin{tabular}{lccc}
\toprule
\textbf{Metric} & \textbf{F} & \textbf{S} & \textbf{F + S} \\
\midrule
Executable (/1000)       & 95.1\% & 83.2\% & 91.6\% \\
Semantic Richness (/1000) & 82.4\% & 91.9\% & 89.7\% \\
\midrule
Semantic Consistent (/100) & 62.0\% & 76.0\% & 71.0\% \\
\bottomrule
\end{tabular}
}
\end{table}

As shown in Table~\ref{tab:stage_strategy}, the three strategies exhibit different strengths. The feedback-driven strategy (F) achieves the highest executability (95.1\%), but its semantic quality is weaker, with lower semantic richness (82.4\%) and lower semantic consistency under manual inspection (62.0\%). In contrast, the semantic-constrained strategy (S) performs best in preserving semantics about the raw PoCs, achieving the highest semantic richness (91.9\%) and semantic consistency (76.0\%), but its executability is the lowest (83.2\%). The strategy used in \BugForge{} (F + S) achieves more practical results for automated testing: it preserves relatively high executability (91.6\%) while also maintaining strong semantic richness (89.7\%) and reasonable semantic consistency (71.0\%), which means it can generate more high-quality test cases compared to the previous two strategies. These results indicate that \BugForge{} is able to make problematic test cases executable while preserving semantic richness as much as possible.

\textbf{Extensibility of \BugForge{}.}
The extensibility of \BugForge{} mainly comes from its design, which separates a shared adaptation engine from DBMS-specific execution backends. In general, this design provides a reasonable basis for extensibility, as the core logic for feedback-driven adaptation and semantic-constrained adaptation from raw PoCs to high-quality test cases is largely independent of any particular DBMS.  In particular, anchor capturing mainly relies on bug reports and raw PoCs, while the feedback generation depends on an executable framework and the feedback signals returned by the target system. Therefore, extending \BugForge{} to a new DBMS does not necessarily require redesigning the entire workflow. In most cases, it only requires implementing system-specific components, such as execution interfaces, dialect-aware guidance, error normalization, and environment reset or cleanup mechanisms.  This also suggests that, as long as the target bug reports can be collected and the target system can return usable execution feedback, the framework can be applied to new DBMS easily. However, applying \BugForge{} to systems with limited documentation or unavailable source code, highly specialized SQL dialects, or complex deployment environments may require additional engineering effort.

\section{Related Work}
\textbf{Bug Repository Construction.}
Bug repositories are a foundational resource for software maintenance and research, containing vast amounts of critical defect information.
Anvik et al.~\cite{10.1145/1117696.1117704} characterize and utilize an open bug repository to analyze the problems of bug triage and duplicate bug detection.
Based on historical reports, Zhang et al.~\cite{10.1145/2695664.2695872} predict the severity of new bugs by binding keywords from bug reports to their corresponding severity levels.
To construct high-quality bug repositories, Williams et al.~\cite{1463230} collect bugs from open-source projects by automatically mining their commit history to identify common patterns for specific types of bug fixes.
BugBuilder~\cite{jiang2022bugbuilder} builds the bug repository by separating pure bug-fixing patches from version control history, thereby ensuring the reproducibility of a subset of the bugs.

However, converting raw DBMS bug reports into high-quality, reproducible PoCs remains challenging.
To bridge this gap, \BugForge introduces an automated bug repository construction method. 
It leverages heuristic analysis and LLM to iteratively fix and complete PoCs, establishing a high-quality repository for automated testing. 

\textbf{Bug Report Based Testing.}
Many works apply information from bug reports to improve the effectiveness of software testing.
BRMINER~\cite{ouedraogo2025enriching} combines traditional extraction techniques with LLM filtering to extract highly relevant test inputs from bug reports, injecting them into existing test generation tools to enhance their bug detection capabilities and code coverage.
LERE~\cite{10.1145/3551349.3556894} automatically extracts real-world test programs from compiler bug reports, and combined with differential testing, can effectively discover new compiler bugs.
PerfLearner~\cite{10.1145/3238147.3238204} automatically generates test frames to guide performance testing by extracting execution commands and key input parameters from performance-related bug reports.
YAKUSU~\cite{10.1145/3213846.3213869} combines program analysis and Natural Language Processing (NLP) to automatically translate natural language bug reports from mobile applications into executable UI test cases.
IBIR~\cite{10.1145/3542946} injects more realistic faults, similar to those found in the real world, by reversing the code transformation templates from a bug-report-driven automated program repair system to enhance software testing capabilities.

However, applying these methods to DBMSs is challenging due to varying SQL dialects and numerous version iterations. 
To address this, \BugForge employs a feedback-driven mechanism to perform tailored analysis, enabling the effective handling of complex bug reports across different dialects.

\textbf{DBMS Testing.}
Mainstream techniques for database testing include fuzzing~\cite{sqlsmith,squirrel,griffin,lego,sqlright}, regression testing~\cite{10.1145/372202.372342, lo2010framework, rogstad2013test, apollo} and cross-DBMS testing~\cite{10.1145/3698829, sedar, gavriilidis2023xdb}.
Fuzzing is applied to test DBMS and discover new bugs. SQLsmith~\cite{sqlsmith} generates test cases from predefined abstract syntax tree templates. 
LEGO~\cite{lego} tests databases by compiling and recombining sequences of SQL queries to generate new ones. 
Squirrel~\cite{squirrel} and Griffin~\cite{griffin} apply mutation strategies to randomly alter initial seeds, generating diverse test cases to uncover DBMS vulnerabilities. 
However, their effectiveness is highly dependent on the quality of the initial seeds.
Regression testing is employed to identify bugs between different database releases.
Apollo~\cite{apollo} automatically detects performance regression bugs in database systems using feedback-driven techniques and query minimization.
Cross-DBMS testing leverages test cases from different database systems to discover new bugs or verify the correctness of SQL queries. 
Sedar~\cite{sedar} employs this technique by transferring test cases between databases to construct a high-quality seed corpus.
Zhong et al.~\cite{10.1145/3698829} uncover new bugs by integrating and reusing test suites from various open-source DBMSs for cross-testing.

Unlike existing tools that rely on inefficient random inputs, \BugForge systematically constructs a repository from real-world bug reports. This approach provides high-quality test cases that dramatically improve DBMS testing.

\section{Conclusion}
In this paper, we propose \BugForge, a framework that constructs and utilizes DBMS bug repositories to enhance DBMS testing. 
\BugForge collects reports, mines bug information with syntax-aware processing and input-adaptive LLM framework, and generates high-quality test cases through semantic-guided adaptation, creating a unified resource for DBMS testing.
Totally, \BugForge generated a repository with \totalreport{} reports and uncovered \totalBugs previously unknown bugs through testing utilization.
In the future, we plan to extend \BugForge{} with more DBMSs and leverage it to analyze bug patterns for discovering new bugs and improving the reliability of DBMSs.





\bibliographystyle{IEEEtran}
\bibliography{ref}


\end{document}